\documentclass[quantumrep,article,submit,moreauthors]{Definitions/mdpi}

\renewcommand{\linenumbers}[1][]{\nolinenumbers}

\providecommand{\showchanges}{0}
\definecolor{revisionOld}{HTML}{D62728} 
\definecolor{revisionNew}{HTML}{2CA02C} 
\DeclareRobustCommand{\deleted}[1]{\ifnum\showchanges=1 {\color{revisionOld}\sout{#1}}\fi}
\DeclareRobustCommand{\added}[1]{\ifnum\showchanges=1 {\color{revisionNew}\uline{#1}}\else#1\fi}
\DeclareRobustCommand{\revised}[2]{\ifnum\showchanges=1\deleted{#1}\space\fi\added{#2}}
\newcommand{\oldmath}[1]{{\color{revisionOld}\sout{\mbox{$\displaystyle #1$}}}}
\newcommand{\newmath}[1]{\ifnum\showchanges=1{\color{revisionNew}\uline{\mbox{$\displaystyle #1$}}}\else#1\fi}

\firstpage{1}
\pubvolume{1}
\issuenum{1}
\articlenumber{0}
\pubyear{2026}
\copyrightyear{2026}
\datereceived{ }
\daterevised{ }
\dateaccepted{ }
\datepublished{ }
\Title{\texorpdfstring{Did the Gran Sasso Experiment Rule Out the Diósi-Penrose Theory? Musing on the 2021 Result and the Present Experimental Status}{Did the Gran Sasso Experiment Rule Out the Diósi-Penrose Theory? Musing on the 2021 Result and the Present Experimental Status}}

\Author{Catalina Curceanu $^{1,2}$\orcidA{}, Nicola Bortolotti $^{3,1}$\orcidB{}, Lajos Diósi $^{4,5}$\orcidC{}, Simone Manti $^{1,}$*\orcidD{} and Kristian Piscicchia $^{6,1}$\orcidE{}}
\AuthorNames{Catalina Curceanu, Nicola Bortolotti, Lajos Diósi, Simone Manti and Kristian Piscicchia}
\address{%
$^{1}$ \quad Laboratori Nazionali di Frascati INFN, Via Enrico Fermi 54, 00044, Frascati, Italy\\
$^{2}$ \quad IFIN-HH, Institutul National pentru Fizica si Inginerie Nucleara Horia Hulubei, 30 Reactorului, 077125, Magurele, Romania\\
$^{3}$ \quad Faculty of Physics, Astronomy and Applied Computer Science, Jagiellonian University, S. {\L}ojasiewicza 11, 30-348, Kraków, Poland\\
$^{4}$ \quad Wigner Research Center for Physics, H-1525 Budapest 114, P.O.Box 49, Hungary\\
$^{5}$ \quad Eötvös Loránd University, H-1117 Budapest, Pázmány Péter stny, 1/A, Hungary\\
$^{6}$ \quad Centro Ricerche Enrico Fermi, Museo Storico della Fisica e Centro Studi e Ricerche ``Enrico Fermi'', Via Panisperna 89/A, 00184 Roma, Italy}
\corres{Correspondence: \href{mailto:simone.manti@lnf.infn.it}{Simone.Manti@lnf.infn.it} (S.M.)}
\abstract{\texorpdfstring{In 2021, a low-background experiment at the Gran Sasso National Laboratory searched for spontaneous radiation predicted by the standard Markovian Diósi-Penrose (DP) dynamics and obtained the new bound \(R_0>5.4\times10^{-11}\,\mathrm m\) at 95\% probability for the short-length cutoff parameter. For germanium at the experimental temperature, the parameter-free prescription identifying \(R_0\) with the spatial quantum spread of a nucleus gives \(R_0\simeq5\times10^{-12}\,\mathrm m\). That particular combination - Diósi's stochastic dynamics and the Penrose-motivated mass-density resolution - was therefore excluded. Subsequent searches strengthened the cutoff bound. The Majorana Demonstrator improved it in 2022, and XENONnT has now obtained the world-leading limits \(R_0>4.9\times10^{-10}\,\mathrm m\) at 90\% confidence level and \(R_0>4.5\times10^{-10}\,\mathrm m\) at 95\% confidence level. These advances reinforce the 2021 exclusion but do not alter its logical scope: they \emph{parametrically} \emph{constrain} the Markovian, effectively white-noise DP dynamics, not Penrose's (and Diósi's) lifetime criterion independently of a dynamics, and not all colored-noise, dissipative, or otherwise modified gravity-related spontaneous collapse theories. We reconstruct what Gran Sasso did and did not rule out and place that conclusion in the present experimental context.}{In 2021, a low-background experiment at the Gran Sasso National Laboratory searched for spontaneous radiation predicted by the standard Markovian Diósi-Penrose (DP) dynamics and obtained the new bound R₀ > 5.4 × 10⁻¹¹ m at 95\% probability for the short-length cutoff parameter. For germanium at the experimental temperature, the parameter-free prescription identifying R₀ with the spatial quantum spread of a nucleus gives R₀ ≃ 5 × 10⁻¹² m. That particular combination - Diósi's stochastic dynamics and the Penrose-motivated mass-density resolution - was therefore excluded. Subsequent searches strengthened the cutoff bound. The Majorana Demonstrator improved it in 2022, and XENONnT has now obtained the world-leading limits R₀ > 4.9 × 10⁻¹⁰ m at 90\% confidence level and R₀ > 4.5 × 10⁻¹⁰ m at 95\% confidence level. These advances reinforce the 2021 exclusion but do not alter its logical scope: they parametrically constrain the Markovian, effectively white-noise DP dynamics, not Penrose's (and Diósi’s) lifetime criterion independently of a dynamics, and not all colored-noise, dissipative, or otherwise modified gravity-related spontaneous collapse theories. We reconstruct what Gran Sasso did and did not rule out and place that conclusion in the present experimental context.}}
\keyword{spontaneous wave-function collapse; Diósi-Penrose model; gravity-related collapse; spontaneous radiation; Gran Sasso; mass-density resolution}

\AtBeginDocument{\hypersetup{hidelinks}}
\begin{document}

\section{The issue}\label{the-issue}

In 2021, we reported an underground search for spontaneous radiation associated with gravity-related wave-function collapse \cite{Donadi2021}. The experiment improved the lower limit on the spatial resolution \(R_0\) of the mass density entering the standard DP master equation by about three orders of magnitude over previous laboratory bounds. Its abstract stated that the result ``rules out the natural parameter-free version'' of the model. The body of the paper identified the assumptions and the numerical prescription to which that statement referred.

In later discussions, however, the qualifier was sometimes lost. ``The parameter-free version of the DP model is ruled out'' became, in compressed form, ``DP is ruled out'', or even ``gravity-related collapse is ruled out''. These statements are not equivalent. The name DP is used for a family of related, but logically distinct, ingredients: a gravitational estimate for the instability of a mass-density superposition; a stochastic dynamical equation producing localization; a prescription for resolving the mass density at short distances; and \revised{assumptions about the temporal spectrum}{assumptions about the frequency spectrum} of the stochastic field. An experiment can constrain only a specified combination of these ingredients.

The purpose of this note is therefore precise. We do not revise the data analysis of Ref. \cite{Donadi2021}, and we do not weaken its actual conclusion. We spell out what was tested, what was excluded, and what remains open. We then follow the experimental development through the Majorana Demonstrator and XENONnT. The newer bounds strengthen the experimental case while making the same conceptual discipline even more important: the strength of a result lies not in the breadth of its slogan, but in the accuracy of the proposition it falsifies.

\section{DP spontaneous wave-function collapse}\label{dp-spontaneous-wave-function-collapse}

\subsection{The gravitational collapse scale}\label{the-gravitational-collapse-scale}

Quantum mechanics of massive objects is in tension with a sharp classical space-time structure. Penrose showed that the time-translation generator is ill-defined for a superposition of different space-time geometries \cite{Penrose1996,Penrose2014}, while Diósi reached an analogous conclusion in the non-relativistic limit from the impossibility of arbitrarily sharp measurements of the Newtonian potential \cite{Diosi1987,Diosi1989}. Both approaches associate the resulting uncertainty of the gravitational structure with the spontaneous collapse of spatial superpositions of sufficiently large masses\added{~\mbox{\cite{Diosi2022}}}. For the collapse rate, they arrive at the following non-relativistic expression.

Consider two alternative mass-density distributions, \(\rho_1(\mathbf r)\) and \(\rho_2(\mathbf r)\), belonging to the branches of a spatial quantum superposition \(\lvert\rho_1\rangle+\lvert\rho_2\rangle\). Their difference \(\delta\rho(\mathbf r)=\rho_1(\mathbf r)-\rho_2(\mathbf r)\) defines a positive gravitational self-energy, written in the convention adopted throughout this article as

\begin{equation}\label{eq:self-energy}
E_{\mathrm{DP}}=\frac{G}{2}\int\!\mathrm d^3\mathbf r\int\!\mathrm d^3\mathbf r^{\prime}\,\frac{\delta\rho(\mathbf r)\,\delta\rho(\mathbf r^{\prime})}{\lvert\mathbf r-\mathbf r^{\prime}\rvert}\, .
\end{equation}

The superposition \(\lvert\rho_1\rangle+\lvert\rho_2\rangle\) collapses spontaneously into one of its two components at the following characteristic collapse time:

\begin{equation}\label{eq:lifetime}
\tau_{\mathrm{DP}}=\frac{\hbar}{E_{\mathrm{DP}}}\, .
\end{equation}

The conjectured gravity-related spontaneous collapse of massive superpositions and the above simple collapse rate mean the DP model \emph{in the narrow sense}.

\subsection[Why is finite spatial resolution inevitable?]{\revised{Why a finite spatial resolution is unavoidable}{Why is finite spatial resolution inevitable?}}\label{why-a-finite-spatial-resolution-is-unavoidable}

If the mass density of a constituent particle is treated as exactly point-like, Eq. \eqref{eq:self-energy} contains an ultraviolet divergence. The gravitational self-energy becomes infinite and Eq. \eqref{eq:lifetime} would predict instantaneous collapse. A finite spatial resolution is therefore not optional: point-like masses must be replaced by smeared mass distributions. We denote the characteristic width of the smearing by \(R_0\). The dependence is physically simple. A smaller \(R_0\) resolves the mass density more sharply, increases \(E_{\mathrm{DP}}\), and produces faster collapse.

There are two conceptually different ways to specify the smearing\deleted{~\mbox{\cite{Ghirardi1990,Diosi2013,Bahrami2014}}}. One may introduce \(R_0\) as a universal phenomenological short-length cutoff to be bounded by experiment\added{~\mbox{\cite{Ghirardi1990,Diosi2013,Bahrami2014}}}. The theory is then a one-parameter \(R_0\)-dependent family. Alternatively, following Penrose\added{~\mbox{\cite{Penrose2014}}}, one may take the actual positional spread of the massive constituents (typically the nuclei) in the material. In a crystal this spread is inferred from the Debye-Waller factor. No new universal constant is then chosen: for a specified material and temperature the resolution is fixed by known quantum and thermal motion. This was called in Ref. \cite{Donadi2021} the natural parameter-free version of the DP model.

\subsection{In what do Diósi and Penrose disagree?}\label{in-what-do-diuxf3si-and-penrose-disagree}

A collapse time \eqref{eq:lifetime} is not yet a collapse dynamics. The standard collapse of superpositions, known from textbooks, is random. Adopting stochasticity, Diósi constructed the simplest Markovian quantum dynamics of universal spontaneous collapse that yields the DP collapse time \eqref{eq:lifetime}. The density operator obeys a Lindblad master equation whose collapse term is a double commutator of the smeared mass-density operator with the Newton kernel \cite{Diosi1987,Diosi1989}. The equation localizes massive spatial superpositions, but it also drives a tiny universal momentum diffusion of all microscopic constituents. Both authors proposed the lifetime estimate in Eq. \eqref{eq:lifetime}, but Penrose did not supply a dynamics. No concrete evolution law, and hence no unique set of secondary experimental signatures, presently follows from his argument alone.

For a long time, a direct experimental test of the DP criterion appeared extremely difficult. Marshall, Simon, Penrose, and Bouwmeester proposed an optomechanical experiment in which a small mirror would be placed in a spatial superposition and the loss of interference monitored \cite{Marshall2003}. The required isolation from environmental decoherence, in particular thermal noise, remains experimentally demanding. A direct test of the DP model in the narrow sense therefore still awaits improved laboratory techniques. On the other hand, the collapse rule \eqref{eq:lifetime} becomes indirectly testable if one adopts Diósi's underlying stochastic dynamics. The Gran Sasso experiment did not create a macroscopic spatial superposition to observe its decay. It tested one unavoidable side effect of the standard Diósi dynamics: noise-driven charged constituents radiate. Consequently, the result is conditional on that dynamics. Combining Diósi's stochastic dynamics with Penrose's prescription for the mass-density resolution is natural and testable, but it is a combination. Excluding that combination does not amount to excluding every possible dynamics compatible with Penrose's lifetime estimate.

\section{The Gran Sasso experiment}\label{the-gran-sasso-experiment}

\subsection{Concept}\label{concept}

In the standard DP master equation, the localization noise produces Brownian-like diffusion even when the system is already well localized. Charged particles accelerated by this diffusion emit radiation. In the energy regime used in Ref. \cite{Donadi2021}, nuclear protons radiate coherently within each nucleus, whereas different nuclei contribute incoherently. \revised{Considered that electrons are relativistic in the analyzed energy range, the predicted spectrum for a material component \(i\) has the form}{The incoherent electronic contribution scales linearly with the atomic number \mbox{\(N\)}, whereas coherent nuclear emission scales as \mbox{\(N^2\)}. We therefore neglect the smaller electronic contribution. The predicted radiation spectrum has the form}

\begin{equation}\label{eq:spectrum}
\begin{gathered}
\ifnum\showchanges=1\oldmath{\frac{\mathrm d\Gamma_i}{\mathrm d\omega}=\frac{2}{3}\,\frac{G e^2 N_i^2 N_{a,i}}{\pi^{3/2}\varepsilon_0 c^3 R_0^3\omega}\, .}\\[8pt]\fi
\newmath{\frac{\mathrm d\Gamma}{\mathrm d\omega}=\frac{2}{3}\,\frac{G e^2 N^2 N_a}{\pi^{3/2}\varepsilon_0 c^3 R_0^3\omega}\, .}
\end{gathered}
\end{equation}

\revised{Here \(N_i\) is the atomic number, \(N_{a,i}\) the number of atoms, and the remaining symbols have their usual meanings.}{Here \mbox{\(N=32\)} is the atomic number of germanium, \mbox{\(N_a\)} the number of germanium atoms, and the remaining symbols have their usual meanings.} Equation \eqref{eq:spectrum} displays the experimentally decisive features: the intensity grows with the amount of material and with the square of atomic number, falls as \(1/\omega\), and is extremely sensitive to \(R_0\) through \(R_0^{-3}\).

This prediction assumes the standard Markovian DP dynamics. \revised{Equivalently, the stochastic field}{Accordingly, the stochastic field} has a sufficiently broad, effectively white spectrum over the frequencies contributing to the measured MeV photons. A colored-noise non-Markovian modification can suppress high-frequency emission while retaining low-frequency localization. Such a modification is a different dynamics and is not constrained by Eq. \eqref{eq:spectrum} in the same way \cite{AdlerBassi2007,AdlerBassi2008,VinanteUlbricht2021}.

\subsection{Apparatus and analysis}\label{apparatus-and-analysis}

The measurement was performed in the low-background environment of the INFN Gran Sasso National Laboratory, where the cosmic-ray flux is reduced by nearly six orders of magnitude. A high-purity germanium detector was surrounded by copper and lead shielding, including 62 kg of electropolished oxygen-free high-conductivity copper in Marinelli geometry. The acquisition corresponded to about 62 days. The detector and its environment were modeled with a validated GEANT4 Monte Carlo simulation to estimate known radioactive backgrounds and the detection efficiency for photons emitted in each component.

The statistical analysis used the interval 1.0-3.8 MeV, where the measured and simulated background spectra agreed sufficiently well and the nonrelativistic nuclear calculation was applicable. In that region 576 events were measured and 506 background events were predicted. A possible DP signal, propagated through the energy-dependent efficiencies of all relevant components, scales as \(z_s(R_0)=a/R_0^3\), with \(a\simeq 1.8\times10^{-29}\,\mathrm m^3\) for the apparatus and exposure \cite{Donadi2021}.

\subsection{Result}\label{result}

No statistically compelling excess attributable to DP radiation was found. The posterior probability distribution yielded

\begin{equation}\label{eq:gran-sasso}
R_0>0.54\times10^{-10}\,\mathrm m\qquad\text{(95\% probability).}
\end{equation}

For the germanium crystal at liquid-nitrogen temperature, the Debye-Waller factor gives a root-mean-square nuclear displacement

\begin{equation}\label{eq:spread}
R_{\mathrm P}=\sqrt{\langle u^2\rangle}=\sqrt{\frac{B}{8\pi^2}}\simeq0.05\times10^{-10}\,\mathrm m\, .
\end{equation}

Thus the lower experimental limit is about eleven times larger than the parameter-free Penrose-motivated value for germanium. Within the assumptions used to derive Eq. \eqref{eq:spectrum}, this is a clear exclusion, not a marginal tension.

\section{Experimental progress after Gran Sasso}\label{experimental-progress-after-gran-sasso}

\subsection{The Majorana Demonstrator}\label{the-majorana-demonstrator}

The Gran Sasso number did not remain the strongest bound. In 2022 the Majorana Collaboration reported a search for spontaneous radiation using 37.5 kg year of enriched-germanium exposure in the 19-100 keV interval \cite{Arnquist2022}. An erratum subsequently corrected the quoted DP result to \(R_{\mathrm{DP}}>(2.54\pm0.03)\times10^{-10}\,\mathrm m\) at 95\% confidence level in the normalization used by Majorana \cite{Arnquist2023Erratum}. Once converted to the Diósi convention used below, this corresponds to approximately \(8.7\times10^{-11}\,\mathrm m\). Majorana thus provided an independent low-background confirmation that the cutoff in the standard Markovian DP dynamics had to be larger than the value inferred at Gran Sasso.

This step was important for two reasons. Experimentally, it exploited a different apparatus, exposure, energy interval, and background analysis. Conceptually, it showed that the Gran Sasso exclusion was not a peculiarity of one detector or one statistical treatment. At the same time, the interpretation remained conditional on the radiation predicted by a white, Markovian collapse noise.

\subsection{XENONnT}\label{xenonnt}

The current strongest direct radiation bound comes from the first science run of the XENONnT dark-matter detector \cite{Aprile2026}. XENONnT used 1.16 tonne year of low-energy electronic-recoil data in liquid xenon over the interval 1-140 keV. In this regime the photon wavelength is comparable with atomic dimensions. The high-energy approximation in which nuclear protons emit coherently and electrons contribute independently is no longer sufficient. The analysis therefore used the low-energy emission theory developed in Ref. \cite{Piscicchia2024}, including interference and cancellation between the oppositely charged nuclear and electronic emitters.

No significant DP-like excess was observed; the local discovery significance was 0.2 standard deviations. XENONnT obtained

\begin{equation}\label{eq:xenon}
\begin{aligned}R_0&>4.9\times10^{-10}\,\mathrm m&&\text{(90\% C.L.),}\\R_0&>4.5\times10^{-10}\,\mathrm m&&\text{(95\% C.L.).}\end{aligned}
\end{equation}

At 95\% confidence, the XENONnT limit is about 8.3 times the Gran Sasso 2021 bound.

\subsection{Convention and comparability of the limits}\label{convention-and-comparability-of-the-limits}

Numerical comparisons require a convention. The normalization of the gravitational self-energy and hence of the DP master equation is not uniform in the literature; the Diósi and Penrose conventions differ by a factor \(8\pi\) in the emission rate used by XENONnT \cite{Aprile2026}. Since the leading emission rate scales as \(R_0^{-3}\), the corresponding numerical cutoff changes by the cube root \((8\pi)^{1/3}\). Throughout this note, experimental values are compared in the Diósi convention used for the stochastic dynamics and for the XENONnT DP limit. This conversion explains why XENONnT's approximately factor-five improvement over Majorana cannot be inferred by directly dividing the two cutoff values as originally printed.

\begin{figure}[t]
\centering
\makebox[\linewidth][c]{\includegraphics[width=5.47in]{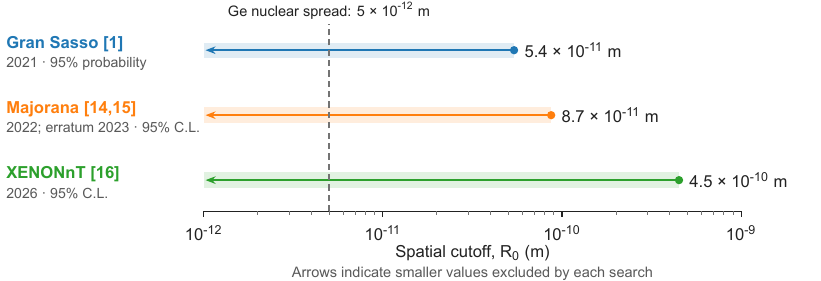}}
\caption{\added{Lower bounds on the spatial cutoff \mbox{\(R_0\)} for the standard Markovian, white-noise DP dynamics, expressed in the Diósi convention. Filled circles show Gran Sasso (2021; 95\% posterior probability), Majorana (2022, corrected in 2023; 95\% confidence level) and XENONnT (2026; 95\% confidence level)~\mbox{\cite{Donadi2021,Arnquist2022,Arnquist2023Erratum,Aprile2026}}.}}
\label{fig:r0-limits}
\end{figure}

The progression Gran Sasso-Majorana-XENONnT is nevertheless physically coherent: all three are searches for radiation generated by a Markovian DP stochastic dynamics. They differ in material, energy window, exposure, background model, and treatment of atomic structure. XENONnT is not merely a rescaling of the Gran Sasso calculation; at low photon energies, atomic charge correlations change both the normalization and the spectral shape. A recent work has placed these effects in a general radial-distribution framework for germanium and xenon \cite{Manti2026}, providing a route toward still more robust material-dependent analyses. \added{In Figure~\mbox{\ref{fig:r0-limits}} the three radiation bounds in this common convention are compared.}

\subsection{What the newer bounds change - and what they do not}\label{what-the-newer-bounds-change---and-what-they-do-not}

The recent experimental bounds make the exclusion of the Penrose-motivated resolution \eqref{eq:spread} still more remote from the experimental boundary. They do not change the logical lesson of 2021. The predicted photons follow from a specific Markovian dynamics where the spectrum of the underlying stochastic field is flat. Majorana and XENONnT strengthen the constraint on that dynamics. XENONnT itself explicitly formulates its DP result for the Markovian, white-noise theory and notes that wider parameter space remains for colored or dissipative extensions \cite{Aprile2026}.

\section{What was and what was not ruled out}\label{what-was-and-what-was-not-ruled-out}

\subsection{What was ruled out}\label{what-was-ruled-out}

The Gran Sasso experiment ruled out the regime \(R_0<5.4\times10^{-11}\) for the spatial resolution (cutoff) parameter of current DP model, i.e. the DP model in narrow sense, embedded in Diósi's microscopic collapse dynamics.

This result excludes the parameter-free version of the DP model, which assumes Diósi's collapse dynamics and identifies the mass-density resolution with the measured positional spread of nuclei in germanium, \(R_0=R_{\mathrm P}\simeq5\times10^{-12}\,\mathrm m\). This is the model referred to in Ref. \cite{Donadi2021} as the natural parameter-free version of the DP model. More compactly, assuming Diósi's standard stochastic dynamics, the Penrose-motivated parameter-free prescription is ruled out.

\subsection{What was not ruled out}\label{what-was-not-ruled-out}

The cutoff-dependent one-parameter DP model was not ruled out. Equation \eqref{eq:gran-sasso} was a lower bound, not a contradiction for every \(R_0\); the later experiments shifted that lower bound upward. At present, standard Markovian DP dynamics with \(R_0\) above the XENONnT 95\% confidence limit, \(4.5\times10^{-10}\,\mathrm m\), is not excluded by these radiation searches. Whether such a value produces sufficiently effective macroscopic collapse, and whether it is theoretically compelling, are separate questions from whether it is experimentally excluded.

Penrose's (and Diósi's) gravitational lifetime criterion \eqref{eq:lifetime} in itself, i.e., the DP model in narrow sense, was not tested at all. Therefore neither its cutoff parameter-dependent nor the parameter-free version was ruled out. The experiment inferred radiation from Diósi's stochastic dynamics; \revised{Penrose's argument by itself does not uniquely predict that radiation spectrum}{Penrose's argument by itself does not predict any radiation}. The data therefore do not exclude every future dynamical completion that might reproduce \revised{\(\tau\simeq\hbar/E_{\mathrm{DP}}\)}{\mbox{\(\tau_{\mathrm{DP}}\simeq\hbar/E_{\mathrm{DP}}\)}}.

Furthermore, colored-noise non-Markovian, dissipative, relativistic, or otherwise modified gravity-related spontaneous collapse dynamics were not generically ruled out. In particular, suppressing the noise spectrum at high frequency can weaken the MeV photon bound. Such modifications must, however, be specified quantitatively and confronted with other experiments at lower frequencies. They cannot be invoked merely as an unspecified escape from one result.

Most importantly, the experiment did not rule out the broad idea that gravity may play a role in wave-function collapse. An experiment falsifies predictions, not motivations. The result tells us that one especially economical free-parameter realization of that motivation does not describe nature.

\section{Discussion and conclusion}\label{discussion-and-conclusion}

The terminology ``Diósi-Penrose model'' has been useful because it recognizes that two independent lines of thought led to \revised{the same gravitational self-energy scale}{the same gravity-related collapse rate}. It can also obscure a difference that becomes decisive in experiment. Diósi's proposal is a dynamics; Penrose's proposal is, at present, a criterion for an instability time. The short-distance resolution can be a free cutoff or a material-dependent quantum width. \revised{The temporal noise can be Markovian or colored.}{The noise can be white or colored.} These choices lead to different experimental predictions and must not be merged after the fact into a single yes-or-no label.

The Gran Sasso result remains historically and conceptually strong. It improved the direct photon-emission bound on \(R_0\) by roughly three orders of magnitude and excluded, by a wide margin, the parameter-free prediction obtained from the nuclear wave-function spread in germanium. Majorana and XENONnT subsequently pushed the cutoff-dependent Markovian theory further, with the present 95\% confidence lower limit reaching \(4.5\times10^{-10}\,\mathrm m\). The proper conclusion is not that ``nothing was ruled out'', but equally not that ``DP was ruled out''. What was first ruled out at Gran Sasso was a precise, natural, and historically important realization of DP collapse; what later experiments have done is enlarge the excluded domain of its cutoff-dependent Markovian continuation.

The remaining theories do not receive immunity from experiment. A free \(R_0\) is now subject to the XENONnT bound in the standard Markovian theory; a colored spectrum must specify its correlation time and confront force-noise, heating, interferometric, and radiation data across many frequency scales; a new dynamical completion of Penrose's criterion must yield quantitative predictions. Precision of language and precision of theory are allied here. The experimental sequence from Gran Sasso to XENONnT has narrowed the viable theoretical space substantially. Its scientific value is best preserved by stating exactly which regions, and which assumptions, the experiments removed.

\section*{Author Note}

This manuscript is a conceptual clarification of the logical scope of Ref. \cite{Donadi2021} and an update of the subsequent experimental status. It reports no new experimental data.

\authorcontributions{\added{Conceptualization, C.C., N.B., L.D., S.M. and K.P.; writing---original draft preparation, C.C., N.B., L.D., S.M. and K.P.; writing---review and editing, C.C., N.B., L.D., S.M. and K.P. All authors contributed to the scientific discussion and to the interpretation of the theoretical and experimental results reviewed in this article.}}

\funding{\added{This publication was made possible through the support of Grant 62099 from the John Templeton Foundation. The opinions expressed in this publication are those of the authors and do not necessarily reflect the views of the John Templeton Foundation. We acknowledge support from the Foundational Questions Institute and the Fetzer Franklin Fund, a donor-advised fund of the Silicon Valley Community Foundation (Grants No. FQXi-RFP-CPW-2008 and FQXi-MGA-2102), and from the H2020 FET-Open project TEQ (Grant Agreement No. 766900). This article is based upon work from COST Actions CA23130 ``Bridging High and Low Energies in Search of Quantum Gravity'' (BridgeQG), and CA23115 ``Relativistic Quantum Information'' (RQI), supported by COST (European Cooperation in Science and Technology). This work also received support from the European Union's Horizon Europe research and innovation programme under the Marie Skłodowska-Curie Staff Exchanges project NEXUS---``North-south EXchange for Underground Science'' (Grant Agreement No. 101236929). The VIP-2 project was supported by the Austrian Science Fund (FWF) through Grants No. P25529-N20, P30635-N36, and W1252-N27 (Doctoral College ``Particles and Interactions''). K.P. acknowledges support from the Museo Storico della Fisica e Centro Studi e Ricerche ``Enrico Fermi'' through the ``Open Problems in Quantum Mechanics'' project. N.B. acknowledges support from the Narodowe Centrum Nauki (Grant No.\ 2023/50/E/ST9/00576).}

\dataavailability{No new data were created or analyzed in this study. Data and source data associated with the experiment are available with Ref. \cite{Donadi2021}.}

\acknowledgments{\added{This article is based upon work from COST Actions CA23130 ``Bridging High and Low Energies in Search of Quantum Gravity'' (BridgeQG), and CA23115 ``Relativistic Quantum Information'' (RQI)}. We acknowledge the support of the Istituto Nazionale di Fisica Nucleare (INFN) and, in particular, of the INFN Gran Sasso National Laboratory (LNGS), its Director Ezio Previtali, the LNGS staff, and the Low Radioactivity Laboratory for the experimental activities dedicated to the search for spontaneous radiation.}}

\conflictsofinterest{The authors declare no conflict of interest.}

\begin{adjustwidth}{-\extralength}{0cm}
\reftitle{References}
\bibliography{DP_QR_MDPI}
\PublishersNote{}
\end{adjustwidth}
\end{document}